\RequirePackage{lineno}
\documentclass{nature}
\usepackage{xcolor}
\usepackage[T1]{fontenc}

\usepackage{caption}
\usepackage{amsmath}
\usepackage{bm}
\usepackage{braket}
\usepackage{amsmath}
\usepackage{amssymb}
\usepackage{textcomp}
\usepackage[colorlinks]{hyperref}

\usepackage{bbold}
\usepackage{mathdots}

\usepackage{cleveref}

\hypersetup{
	citecolor={blue},
	urlcolor={blue}
}

\title{Chiral Floquet Engineering on Topological Fermions in Chiral Crystals}

\author{Benshu Fan$^{1,2}$, Wenhui Duan$^{1,3,4,*}$, Angel~Rubio$^{2,5,6,\dagger}$, Peizhe Tang$^{7,2,\ddagger}$}

\usepackage{graphicx}
\makeatletter
\let\saved@includegraphics\includegraphics
\AtBeginDocument{\let\includegraphics\saved@includegraphics}
\renewenvironment*{figure}{\@float{figure}}{\end@float}
\makeatother

\begin{document}
	\maketitle
	
	\begin{affiliations}
		\item State Key Laboratory of Low-Dimensional Quantum Physics and Department of Physics, Tsinghua University, Beijing 100084, People's Republic of China
		\item Max Planck Institute for the Structure and Dynamics of Matter, Center for Free Electron Laser Science, 22761 Hamburg, Germany
		\item Institute for Advanced Study, Tsinghua University, Beijing 100084, People's Republic of China
		\item Frontier Science Center for Quantum Information, Beijing 100084, People's Republic of China
		\item Nano-Bio Spectroscopy Group, Departamento de Fisica de Materiales, Universidad del País Vasco UPV/EHU- 20018 San Sebastián, Spain
		\item Center for Computational Quantum Physics (CCQ), The Flatiron Institute, 162 Fifth avenue, New York NY 10010, USA
		\item School of Materials Science and Engineering, Beihang University, Beijing 100191, People's Republic of China\\
		
		$*$e-mail: duanw@tsinghua.edu.cn\\
		$\dagger$e-mail: angel.rubio@mpsd.mpg.de\\
		$\ddagger$e-mail: peizhet@buaa.edu.cn
	\end{affiliations}

	\section*{Abstract}
	\begin{abstract}
		{\bf The interplay of chiralities in light and quantum matter provides an opportunity to design and manipulate chirality-dependent properties in quantum materials. Herein we report the chirality-dependent Floquet engineering on topological fermions with the high Chern number in chiral crystal CoSi via circularly polarized light (CPL) pumping. Intense light pumping does not compromise the gapless nature of topological fermions in CoSi, but displaces the crossing points in momentum space along the direction of light propagation. The Floquet chirality index is proposed to signify the interplay between the chiralities of topological fermion, crystal, and incident light, which determines the amplitudes and directions of light-induced momentum shifts. Regarding the time-reversal symmetry breaking induced by the CPL pumping, momentum shifts of topological fermions result in the birth of transient anomalous Hall signals in non-magnetic CoSi within an ultrafast time scale, which Mid-infrared (IR) pumping and terahertz (THz) Kerr or Faraday probe spectroscopy could experimentally detect. Our findings provide insights into exploring novel applications in optoelectronic devices by leveraging the degree of freedom of chirality in the non-equilibrium regime.}
	\end{abstract}
	
	\newpage
	
	\renewcommand{\thefigure}{\textbf{Fig. \arabic{figure} $\bm{|}$}}
	\setcounter{figure}{0}
	
	\section*{Introduction}
	Chirality, which refers to the misalignment between an object and its mirror image, is a prevalent phenomenon in the natural world\cite{song2016chiral,yang2021chiral,liu2021chirality,fu2022quantum}. In the realm of a quantum material, constrained by chiral space groups (SGs)\cite{chang2018topological}, chiral crystals exhibit a distinct handedness $\eta$ in their lattices due to the absence of mirror, inversion, or roto-inversion symmetries\cite{chang2018topological,li2019chiral,li2022chirality}. For example, within the chiral SG 198 ($P2_13$), the compound family of CoSi hosts left-handed (LH, $\eta=+1$) and right-handed (RH, $\eta=-1$) lattice structures\cite{tang2017multiple,chang2017unconventional}, which are mirror images of each other. Due to the protection of the chiral crystal symmetry, the CoSi family possesses unconventional chiral fermions with higher topological charges in bulk states\cite{tang2017multiple}. Namely, in the LH CoSi without spin-orbit coupling (SOC), spin-1 excitation and double Weyl fermion with topological charges $+2$ and $-2$ emerge at $\Gamma$ and R points around the Fermi level, characterized by the fermion's chirality (defined by the sign of the topological charge\cite{ma2017direct}) $\chi_\Gamma(\eta)=+1$ and $\chi_{\rm{R}}(\eta)=-1$. Upon inclusion of SOC, the chiral fermions at $\Gamma$ and R points become spin-3/2 and double spin-1 fermions\cite{bradlyn2016beyond,hasan2021weyl}, as shown in Fig.~1a. These chiral topological fermions have led to many chirality-dependent intriguing physical phenomena in chiral crystals, a field that has garnered much attention over the past few years. These phenomena include helicoid surface states\cite{schroter2019chiral,rao2019observation,li2019chiral,sanchez2019topological,takane2019observation,schroter2020observation}, circular photogalvanic effect\cite{de2017quantized,flicker2018chiral,rees2020helicity,xu2020optical}, chirality locking charge density wave\cite{li2022chirality}, the interference of chiral quasiparticle\cite{yuan2019quasiparticle,sessi2020handedness}, and other attributes such as unconventional resistivity scaling\cite{lien2023unconventional}, orbital transport\cite{yang2023monopolelike}, exotic excited modes\cite{dutta2022collective,huber2023quantum} and quasi-symmetry-protected topology\cite{guo2022quasi}.
	
	Photon, like various chiral quasiparticles, possesses an optical chirality $\gamma$, referring to the left- and right-circularly polarized light (LCPL, $\gamma=+1$ and RCPL, $\gamma=-1$). Leveraging time-periodic laser fields, Floquet engineering emerges as a potent avenue to realize light-dressed states through virtual-photon absorption or emission processes, thereby enabling an opportunity to engineer electronic structures and topological properties of quantum materials in non-equilibrium\cite{HsiehDemond2017,wang2018theoretical,oka2019floquet,DelaTorre2021,bao2022light}. Especially, with CPL pumping, the time-reversal symmetry (TRS) in the host material is transiently broken, inducing many interesting phenomena that have been observed experimentally, including the dynamical gap opening of Dirac cone on the surface state of topological insulator\cite{WangYH2013,ito2023build}, as well as CPL-induced anomalous Hall conductance in graphene\cite{mciver2020light,sato2019microscopic} and Cd$_3$As$_2$\cite{murotani2023disentangling}. Theoretically, CPL-driven Floquet engineering is proposed in many topological systems, encompassing the momentum shift of chiral Weyl nodes\cite{chan2016chiral}, the splitting of Dirac fermions\cite{hubener2017creating,LiXiaoShi2019}, and the driving of topological phase transitions in semiconductors\cite{lindner2011floquet,Lindner2020,zhan2023floquet,liu2018photoinduced,liu2023floquet,zhu2023floquet} and semimetals\cite{oka2009photovoltaic,Inoue2010,sentef2015theory,yan2016tunable,zhang2016theory,chen2018floquet,trevisan2022bicircular}. While the exploration of the interplay between the chiralities of light and chiral crystals is still lacked, which can provide a new chance to manipulate electronic properties of chiral topological materials.
	
	In this Letter, we report the chiral Floquet engineering on topological fermions in the CoSi family using the Floquet effective $\mathbf{k\cdot p}$ model from the perturbation theory, complemented by the Floquet tight-binding Hamiltonian based on the {\it ab initio} calculations as a benchmark. We show that under CPL pumping, gapless topological fermions with different topological charges undergo momentum shifts along opposite directions (see Fig.~1b) while preserving their topological properties. This phenomenon is distinct from the momentum shifts of Weyl points driven by the light-induced shear phonon mode in WTe$_2$\cite{sie2019ultrafast} and shifts of topological fermions under the linearly polarized light (LPL)\cite{kitayama2021predicted,neufeld2023band}. The sign and magnitude for light-induced momentum shifts at $\Gamma$ and R points are determined by the Floquet chirality index $\Xi_{k=\Gamma, \rm{R}}$, which depends on the light-matter coupling strength and the interplay among the three distinct chiralities, the crystal handedness $\eta$, the chirality $\chi_{k=\Gamma, \rm{R}}(\eta)$ of topological fermions, and the chirality $\gamma$ of CPL. Via analyzing the Lie algebra representations of $\mathfrak{s u}(2)$ for effective Hamiltonians of topological fermions under laser pumping, we provide a comprehensive understanding of chiral Floquet-engineered transient change of electronic structures in CoSi, which could be detected by the Mid-IR pumping and THz Kerr or Faraday probe spectroscopy measurements in an ultrafast time scale\cite{yoshikawa2022light}.
	
	\section*{Results}
	\subsection{$\mathbf{k}\cdot\mathbf{p}$ model analysis}
	~\\
	We start from the effective $\mathbf{k\cdot p}$ model to investigate the evolution of topological fermions in the LH CoSi ($\eta=+1$) under CPL pumping in the framework of the Floquet engineering. For the CoSi compound, electronic structure calculations without SOC reproduce experimental observations faultlessly\cite{rao2019observation,li2019chiral,sanchez2019topological,takane2019observation}, thus in the following we focus on CoSi and do not include the SOC effect (see the Supplementary Note~1). For compounds composed of heavy elements such as AlPt\cite{schroter2019chiral} and PdGa\cite{schroter2020observation,sessi2020handedness}, SOC must be considered seriously and related discussions are shown in Supplementary Note~2 and Note~4. At the $\Gamma$ point, the topological fermion around the Fermi level in the CoSi compound is represented as a spin-1 excitation with $\chi_{\Gamma}(\eta)=+1$\cite{tang2017multiple} and its $\mathbf{k\cdot p}$ Hamiltonian is $\hat{H}_{\Gamma}(\mathbf{k})=\hbar v_{\Gamma}\eta\mathbf{k}\cdot\mathbf{J}$, where $\hbar$ is the reduced Plank constant, $v_{\Gamma}>0$ denotes the Fermi velocity for spin-1 excitation, $\mathbf{k}$ is the momentum and $\mathbf{J}=(J_x,J_y,J_z)$ represents the matrix representation of the spin-1 angular momentum operator. 
	
	Given that CoSi is a non-magnetic material with TRS but lacks inversion symmetry due to its chiral crystal structure, the light-matter interaction will yield distinct phenomena depending on the type of pumping laser used. The LPL can indeed break the inversion symmetry of the material, however, in the case of the chiral crystal CoSi, unlike the previous study on the organic salt with tilted Dirac nodes\cite{kitayama2021predicted}, LPL does not introduce any new changes of symmetry (see the Supplementary Note~3 and Note~4). On the other hand, CPL can break the TRS of CoSi, potentially leading to the emergence of transient anomalous Hall signals\cite{chan2016chiral}. Consequently, we focus exclusively on the CPL in our study. We consider the CPL to be incident along the $x$ direction with a frequency $\Omega$ and a vector potential $\mathbf{A}(t)=A_0(0,\gamma \sin\Omega t, \cos\Omega t)$ where $A_0$ is its amplitude. Then we apply the Peierls substitution to account for the light-matter interaction term and the Floquet theory with the Magnus expansion\cite{Mikami2016Brillouin} to calculate the light-dressed electronic structures for spin-1 excitation, whose Floquet effective $\mathbf{k}\cdot\mathbf{p}$ Hamiltonian is obtained as (see the Supplementary Note~2)
	\begin{equation}
		\hat{H}^{eff}_\Gamma(\mathbf{k})=
		\hbar v_{\Gamma}\eta\left(\mathbf
		{k}\cdot\mathbf{J}-\frac{\gamma v_\Gamma\eta A^2}{2\Omega}J_x\right)\label{eq:gamma_eff}
	\end{equation}
	Herein, we redefine $A=eA_0/\hbar$ as the effective amplitude of the vector potential for simplicity, in which $e$ is the charge of the electron.
	As calculated results illustrated in Fig.~2a, it is evident that the spin-1 excitation with threefold degeneracy in the LH CoSi keeps its gapless nature under the irradiation of the CPL, without the change of its topological charge. Instead, the crossing point shifts along the $x$ direction in momentum space with the magnitude of $\delta_{\Gamma}$, which is given by
	\begin{equation}
		\delta_{\Gamma}=+\frac{\gamma v_\Gamma \eta A^2}{2\Omega}= \frac{\gamma \chi_\Gamma(\eta) A^2}{\Omega}\cdot\frac{v_\Gamma}{2}=\Xi_\Gamma\frac{v_\Gamma}{2}>0
	\end{equation}
	Here, we propose the quantity of $\Xi_\Gamma=\frac{\gamma \chi_\Gamma(\eta) A^2}{\Omega}$ as the Floquet chirality index for the first time, which combines the chirality $\gamma$ of CPL and $\chi_\Gamma(\eta)$ of topological fermion and also depends on the crystal handedness $\eta$. The interplay among these chiralities via Floquet engineering determines the light-induced momentum shifts of spin-1 excitation in CoSi.
	
	For the double Weyl fermion at the R point with opposite topological charge [$\chi_{\rm{R}}(\eta)=-1$], we can demonstrate that its crossing point shifts to an opposite direction in momentum space under the same laser driving, in contrast to the evolution of the spin-1 excitation at the $\Gamma$ point. The effective $\mathbf{k\cdot p}$ Hamiltonian of the double Weyl fermion can be expressed as a direct sum of two decoupled Weyl fermions as $\hat{H}_{\rm{R}}(\mathbf{k})=\hbar v_{\rm{R}}\eta \mathbf{k}\cdot(\bm{\sigma}\oplus\bm{\sigma})$, where $v_{\rm{R}}>0$ represents the Fermi velocity at the R point and $\bm{\sigma}=(\sigma_x, \sigma_y, \sigma_z)$ are
	Pauli matrices. In the same way, we obtain the Floquet effective $\mathbf{k}\cdot\mathbf{p}$ Hamiltonian $\hat{H}^{eff}_{\rm{R}}(\mathbf{k})$ as (see the Supplementary Note~2)
	\begin{equation}
		\begin{aligned}\label{eq:R_eff}
			\hat{H}^{eff}_{\rm{R}}(\mathbf{k})=\hbar v_{\rm{R}} \eta \left(\mathbf{k}\cdot(\bm{\sigma}\oplus\bm{\sigma})+\frac{\gamma v_{\rm{R}} \eta A^2}{\Omega}\sigma_x\oplus\sigma_x\right)
		\end{aligned}
	\end{equation}
	and the shift $\delta_{\rm{R}}$ expresses as
	\begin{equation}
		\delta_{\rm{R}}=-\frac{\gamma v_{\rm{R}} \eta A^2}{\Omega}=\frac{\gamma \chi_{\rm{R}}(\eta) A^2}{\Omega}v_{\rm{R}}=\Xi_{\rm{R}}v_{\rm{R}}<0
	\end{equation}
	which is opposite compared to the light-induced momentum shift of the spin-1 excitation, as shown in Fig.~2b with red dot lines for LCPL ($\gamma=+1$).
	
	Intriguingly, a similar effect of momentum shift is reported for the Weyl fermion upon exposure to CPL\cite{chan2016chiral}. We can take it as an example to understand such an effect for topological fermions in the framework of Lie algebra. The Pauli matrix $\bm{\sigma}$ appearing in the effective Hamiltonian of Weyl fermion ($\hat{H}_{W}\sim\bm{k}\cdot\bm{\sigma}$) is the generator of the Lie algebra $\mathfrak{s u}(2)$. Because of the completeness of the two-dimensional Pauli matrix, the correction term from the Floquet commutator within the Magnus expansion (see the Supplementary Note~2) still can be expressed by the Pauli matrix $\bm{\sigma}$. Thus, under the CPL pumping, there will be a light-induced momentum shift for the gapless Weyl fermion. Such result is in sharp contrast to the dynamic behavior of the Dirac fermion [$\hat{H}_D \sim \mathbf{k}\cdot(\bm{\sigma}\oplus\bm{\sigma^*})$] under the Floquet engineering, which will be converted to a pair of {W}eyl fermions with opposite chirality\cite{hubener2017creating}. We can also understand this phenomenon in the framework of Lie algebra. Because $\bm{\sigma}\oplus\bm{\sigma^*}$ is not the generator of $\mathfrak{s u}(2)$, ultimately the Dirac fermion will undergo a photoinduced momentum splitting and lead to the emergency of Floquet-{W}eyl cones under CPL radiation (see the Supplementary Note~3).

	Within the similar analysis, we found that, for spin-1 and double Weyl fermions, both $\mathbf{J}$ and $\bm{\sigma}\oplus\bm{\sigma}$ can serve as generators of the Lie algebra $\mathfrak{s u}(2)$. We can prove that their generators can still expand the Floquet commutator (see the Supplementary Note~3). Consequently, the Floquet engineering of spin-1 and double Weyl fermions under CPL pumping resembles a similar behavior for Weyl fermions under the same condition. More generally, for the spin-S topological excitation, as long as $\mathbf{S}$ is the generator in the (2S+1)-dimensional irreducible representation of $\mathfrak{s u}(2)$, the crossing point for the Hamiltonian $\hat{H}\sim\mathbf{k}\cdot\mathbf{S}$ will not be gapped under the CPL pumping in the framework of the Floquet engineering, instead, the crossing point will be shifted in momentum space. When we consider the CoSi family compounds with SOC, a spin-3/2 fermion ($\mathbf{k}\cdot\mathbf{S}_{3/2}$) locates at the $\Gamma$ point and a double spin-1 fermion ($\mathbf{k}\cdot\mathbf{J}\oplus\mathbf{k}\cdot\mathbf{J}$) can be found at the R point around the Fermi level. Since $\mathbf{S}_{3/2}$ and $\mathbf{J}$ denote generators in the $4-$ and $3-$dimensional irreducible representations of $\mathfrak{s u}(2)$, light-induced momentum shifts rather than the gap opening occur at crossing points around the Fermi level, and such analysis is consistent with our calculations based on the Floquet effective model (see the Supplementary Note~2).

	\subsection{Tight-binding Hamiltonian analysis}
	~\\
	The discussion shown above is based on perturbation theory and effective $\mathbf{k}\cdot\mathbf{p}$ model that is only valid for high-frequency limit and low energy approximation\cite{oka2019floquet,Mikami2016Brillouin}. To verify the robustness of the phenomena reported here, we apply the Floquet tight-binding calculations for the LH CoSi (see Methods, the Supplementary Note~4 and Note~5) in which the hopping parameters are obtained from \textit{ab initio} calculations. We neglect the influence of dipole transitions, as they do not fundamentally alter the qualitative behavior of the momentum shift in our case, which is supported by previous research on the Dirac semimetal Na$_3$Bi\cite{hubener2017creating}. As shown in Fig.~2a,b, the light-induced momentum shifts can be observed for the gapless topological fermions at $\Gamma$ and R points, which are consistent with results of the Floquet effective $\mathbf{k}\cdot\mathbf{p}$ model. Furthermore, we calculate the distributions of bulk states in the BZ for CoSi under CPL pumping when the energy levels cross the gapless points of spin-1 excitation and double Weyl fermion. The results obtained from the tight-binding calculations are shown in Fig.~2c,d, which are summed over bulk states along the $k_z$ direction and projected on the (001) side surface. Notably, in sharp contrast to the symmetric distribution of bulk states without laser pumping, the light-induced asymmetric feature can be observed in the center of plots (see regions around $\bar{\Gamma}$ in Fig.~2c and $\bar{\rm M}$ in Fig.~2d), attributed to TRS breaking caused by CPL pumping.

	In equilibrium, the non-trivial topology of bulk states results in the emergence of Fermi arcs on the surface, which connect the projected topological fermions with opposite topological charges. So it is natural to explore the evolution of non-equilibrium surface states upon laser pumping. Here upon the LCPL pumping, we calculate the surface band structure projected onto the (001) surface for both LH and RH CoSi, as shown in Fig.~3a. Besides Fermi arcs with the Floquet index $n=0$ which connect the projections of the spin-1 excitation around the $\bar{\Gamma}$ point and the double Weyl fermion around the $\bar{\rm M}$ point, additional replica surface states with Floquet indices of $n=\pm 1$ also can be observed and the energy difference between the neighboring Floquet sidebands equals the pumping photon energy $\hbar\Omega$.
	
	Figure 3a shows the Fermi arc contours on the (001) surface for LH and RH CoSi with different energy levels. Similar to the case in equilibrium\cite{tang2017multiple}, under laser pumping, the Fermi arcs are observed in a large energy window across the whole BZ of the side surface, which connect the projections of bulk states around $\bar{\Gamma}$ and $\bar{\rm M}$ points. Because the LH and RH structures in CoSi are mirror images of each other, correspondingly on the same (001) side surface, the Fermi arcs in LH and RH structures can be converted to each other via the operation of mirror symmetry (see Fig.~3b). Moreover, except for the emergency of additional replica Floquet surface states, the distribution of Floquet Fermi arcs is similar to the case without laser pumping (see the Supplementary Note~1). Upon considering the influence of SOC in CoSi and AlPt, as illustrated in Supplementary Note~4, the degeneracy of surface states is lifted while the connection of the surface states remains consistent with the scenario in the absence of SOC. Such fact indicates that upon the CPL pumping, topological properties of topological fermions in the CoSi family will not be altered although the Floquet states emerge and crossing-point positions are shifted by the pumping laser.

	\section*{Discussion}
	
	In this work, we demonstrate a chiral Floquet engineering of topological fermions in the CoSi family. However, the experimental realization of Floquet engineering is still challenging and only limited to a few material candidates\cite{WangYH2013,mahmood2016selective,ito2023build,Wang2014Stark,sie2017large,Aeschlimann2021Survival,zhou2023pseudospin,zhou2023Floquet,merboldt2024observation,choi2024direct}. One significant obstacle in the formation of Floquet states lies in the competition between the laser pumping and the decoherent scattering for the excited electrons\cite{sato2020floquet}. To avoid the Floquet state's collapse, the pumping laser period ($T=2\pi/\Omega$) should be shorter than the experimental scattering time. For a metallic system, such as the surface state of $\text{Bi}_2\text{Te}_3$, a previous study has shown that the emergency of Floquet-Bloch sidebands can be built up within a few ultrashort optical cycles\cite{ito2023build}. Based on the transport measurement\cite{xu2019crystal}, the scattering time in the CoSi single crystal is estimated as around 131~fs (see the Supplementary Note~6), which provides a lower limit to the pumping laser frequency.
	Consequently, considering these constraints, we expect a Mid-IR pumping laser with photon energy around 100~meV and an electric field intensity as large as $4.4\times10^7$~V/m can be used to excite the Floquet states in CoSi compounds. Additionally, time- and angle-resolved photoemission spectroscopy (TrARPES) is a powerful and direct experimental technique\cite{chen2020observing} to observe Floquet band structures, such as the detection of Dirac surface\cite{WangYH2013,ito2023build} and bulk\cite{merboldt2024observation,choi2024direct} states under the Mid-IR laser pumping (see the Supplementary Note~7 for details). It should be noted that a tunable probe photon energy is essential for the detection process\cite{takane2019observation,rao2019observation}, particularly to track the evolution of topological fermions along the $k_z$ direction when subjected to the Mid-IR laser pumping that propagates along the $z$ direction. Therefore, we propose that TrARPES can be used to probe the light-dressed electronic structures of both bulk and surface states.

	Beyond the direct observation of light-induced Floquet states, we suggest that Mid-IR pumping and THz Kerr or Faraday probe spectroscopy experiments could also be employed to probe the chiral Floquet-engineered electronic structures\cite{yoshikawa2022light}. In the absence of laser pumping, no signal can be detected in non-magnetic CoSi due to the presence of TRS\cite{cheong2023trompe}. However, upon the pumping of CPL to break TRS, in general, both the non-equilibrium electron occupation and the light-induced Berry curvature can contribute to anomalous Hall conductivity (AHC). Based on the previous study in a similar setup\cite{yoshikawa2022light}, we anticipate that the light-induced Berry curvature, which appears in an ultrafast time scale proportional to $\sum_{k\in\{\Gamma, \rm{R}\}}\chi_k(\eta)\cdot\delta_k$\cite{chan2016chiral,yang2011quantum}, could partially contribute to the AHC\cite{yoshikawa2022light,murotani2023disentangling,hirai2023anomalous} (see the Supplementary Note~7 for more detailed analysis). Given that the Kerr and Faraday angles are linearly proportional to the AHC\cite{kahn1969ultraviolet}, we expect that the pump-probe Kerr or Faraday experiments could detect momentum shifts of topological fermions. As proposed in Fig.~4, when we apply the CPL pumping laser on the CoSi thin film with (111) surface, a finite value for Kerr angle $\theta_K$ and Faraday angle $\theta_F$ should be detected in the duration of the pumping laser.

	In summary, applying chiral Floquet engineering on chiral crystals, we demonstrate that the CPL pumping can induce momentum shifts of topological fermions in CoSi and these shifts occur either parallel or antiparallel to the propagation direction of the incident beam depending on the Floquet chirality index $\Xi_k$. Via an analysis of the Lie algebra representation of $\mathfrak{s u}(2)$ for the low-energy Hamiltonians of topological fermions, we could extend our conclusion to other SOC-dominated heavy compounds in CoSi family, such as AlPt and PdGa, and other spin-S fermionic excitations\cite{bradlyn2016beyond} under the chiral Floquet engineering. Furthermore, we propose that light-induced electronic structure changes in CoSi could be detected via TrARPES and Mid-IR pumping and THz Kerr or Faraday probe spectroscopy experiments. Our findings propose leveraging chirality as an adjustable degree of freedom in Floquet engineering, paving new avenues to enable ultrafast switching of material properties and the development of innovative optoelectronic devices, such as chirality logic gates\cite{zhang2022chirality} and chiral optical cavities\cite{hubener2021engineering}.
	
	\begin{methods}
		\subsection{First-principles calculation}
		~\\
		We applied the Vienna \emph{Ab initio} Simulation Package (VASP)\cite{kresse1996efficient} to perform density functional theory (DFT) calculations to investigate electronic ground states of the CoSi and AlPt compounds using the Perdew-Burke-Ernzerhof-type exchange-correlation functional\cite{perdew1996generalized}. Projector augmented wave potentials\cite{kresse1999ultrasoft} were employed with a plane-wave energy cutoff of 300 eV. The Brillouin zone (BZ) of the primitive cell was sampled using a $\Gamma$-centered $9\times9\times9$ k-point grid. The convergence criteria for the electronic self-consistent calculation and force were set to $10^{-7}$ eV and 0.01 eV/\AA, respectively. 
		
		We constructed the tight-binding Hamiltonian $\hat{H}^{TB}(\mathbf{k})$ from the \emph{ab initio} calculations using maximally localized Wannier functions obtained from Wannier90 code\cite{mostofi2008wannier90}. $d$ orbitals of Co (Pt) atoms and $p$ orbitals of Si (Al) atoms were chosen as projected orbitals. The (001) surface projections of bulk states in Figs.~2c and 2d, as well as the surface states in Fig.~3, were obtained by extracting the spectral functions from the imaginary part of the bulk Green's function:
		\begin{equation}
			A_b(k_x, k_y, E_{arc})=-\frac{1}{\pi}\lim_{\epsilon\rightarrow0^+}Im\{Tr[G_b(k_x, k_y, E_{arc}+i\epsilon)]\}
		\end{equation}
		and the surface Green's function:
		\begin{equation}
			A_s(k_x, k_y, E_{arc})=-\frac{1}{\pi}\lim_{\epsilon\rightarrow0^+}Im\{Tr[G_s(k_x, k_y, E_{arc}+i\epsilon)]\}
		\end{equation}
		Here, $E_{arc}$ denotes the energy of the Fermi arc states, and $\epsilon$ is an infinitesimally small broadening parameter for the Fermi arc states. $G_b(k_x, k_y, E_{arc}+i\epsilon)$ and $G_s(k_x, k_y, E_{arc}+i\epsilon)$ are the bulk and surface Green's functions, respectively, which were calculated by the iterative Green’s function method\cite{Sancho1985Highly}. These calculation procedures have been merged into the open-source software package WannierTools\cite{wu2018wanniertools}.
		
		\subsection{Floquet Theory}
		~\\
		The Floquet theory is a powerful mathematical tool that relates the time-periodic Schr\"odinger equation $\hat{H}^{TB}(\mathbf{k}, t)\Phi_\alpha(t) = i\hbar\frac{\partial}{\partial t} \Phi_\alpha(t)$ to a static eigenvalue problem in the energy domain. Herein, we construct the time-periodic tight-binding Hamiltonian or effective model by using the Peierls substitution
		\begin{equation}\label{eq:time-H}
			\hat{H}(\mathbf{k}) \rightarrow \hat{H}(\mathbf{k}, t)=\hat{H}(\mathbf{k}+e\mathbf{A}(t)/\hbar)
		\end{equation}
		where $e$ is the charge of electron, $\hbar$ is the reduced Plank constant and $\mathbf{A}(t)$ is the vector potential of the pumping laser.
		
		For the time-dependent Hamiltonian with one optical cycle $T$ in Eq.~(\ref{eq:time-H}), the Floquet theory guarantees the existence of solutions $\Phi_\alpha(t)$ that can be expressed as 
		\begin{equation}\label{eq:Phi}
			\Phi_\alpha(t)=e^{-iE_\alpha t}u_\alpha(t)
		\end{equation}
		where $E_\alpha$ denotes the Floquet quasienergy and the periodic function $u_\alpha(t)$ satisfies $u_\alpha(t+T) = u_\alpha(t)$. Moreover, $u_\alpha(t)$ can be expanded in a Fourier series with coefficients $u_\alpha^m$ as
		\begin{equation}\label{eq:fourier}
			u_\alpha(t)=\sum_{m=-\infty}^{\infty} e^{-i m \Omega t}u_{\alpha}^{m}
		\end{equation}
		where $\Omega=2\pi/T$ is the frequency of the pumping laser.
		By substituting Eq.~(\ref{eq:Phi}) and Eq.~(\ref{eq:fourier}) into the original Schr\"odinger equation and performing some algebraic manipulations, we can derive an eigenvalue equation for the coefficients $u_\alpha^m$ and the corresponding eigenvalues $E_\alpha$ as
		\begin{equation}
			\sum_{m}\hat{\mathcal{H}}_{nm}(\mathbf{k})u_{\alpha}^{m}=E_{\alpha}u_{\alpha}^{n}
		\end{equation}
		where $\hat{\mathcal{H}}_{nm}(\mathbf{k})=\frac{1}{T}\int_{0}^{T}dt\hat{H}(\mathbf{k}, t) e^{i (n-m) \Omega t}-m\hbar\Omega\delta_{mn}$ is the matrix element of the Floquet tight-binding Hamiltonian or Floquet effective model as $\hat{\mathcal{H}}(\mathbf{k})$ and $m$, $n$ are Floquet indices. We can diagonalize $\hat{\mathcal{H}}(\mathbf{k})$ to obtain Floquet band structures of the CoSi crystal in the main text. 
	\end{methods}
	


	\begin{addendum}
		\item[Data availability]
		~\\
		The data supporting the plots within this article and other findings of this study are available from the corresponding author upon reasonable request.
		
		\item[Code availability]
		~\\
		The code used for this study is not publicly available but may be made available to qualified researchers on reasonable request from the corresponding author.

		\item[Acknowledgements] 
		~\\
		P.T. was supported by the National Natural Science Foundation of China (Grants No. 12234011 and No. 12374053). B.F. and W.D. acknowledge the support of the Basic Science Center Project of NSFC (Grant No. 52388201), Innovation Program for Quantum Science and Technology (Grant No. 2023ZD0300500), and the Beijing Advanced Innovation Center for Future Chip (ICFC). A.R. acknowledges support from the Cluster of Excellence 'CUI: Advanced Imaging of Matter'- EXC 2056 - project ID 390715994, and Grupos Consolidados (IT1453-22). A.R. also acknowledges support from the Max Planck-New York City Center for Non-Equilibrium Quantum Phenomena. The Flatiron Institute is a division of the Simons Foundation.
		
		\item[Author contributions] 
		~\\
		P.T., D.W., and A.R. conceived and supervised the research project. B.F. developed the theoretical description and performed all numerical
		calculations. All authors contributed to the discussions and commented on the manuscript.
		
		\item[Competing interests] 
		~\\
		The authors declare no competing interests.
		
		
		
		
	\end{addendum}

	\subsection{Figure legends}
	\clearpage
	\begin{figure}
		\centering
		\includegraphics[width=0.8\linewidth]{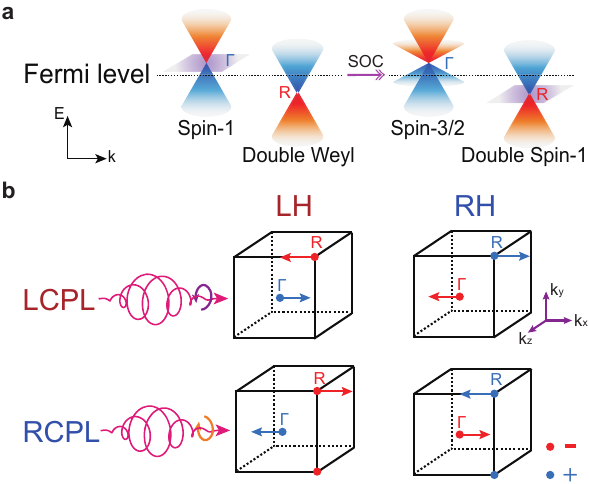}
		\caption{{\bf Multiple types of topological fermions and chiral Floquet engineering in the CoSi compound.} {\bf a} A schema of topological fermions at $\Gamma$ and R points which exhibit spin-1 (spin-3/2) and double Weyl (double spin-1) excitations without (with) SOC around the Fermi level in CoSi. {\bf b} Schema of the left-circularly polarized light (LCPL)- and right-CPL (RCPL)- induced momentum shifts for topological fermions, marked as blue and red arrows, in the LH and RH CoSi crystal via Floquet engineering. The black cube represents the Brillouin zone (BZ) of the CoSi family and the red negative and blue positive signs denote the sign of topological charges.}\label{fig:1}
	\end{figure}
	
	\begin{figure}[h]
		\centering
		\includegraphics[width=0.8\linewidth]{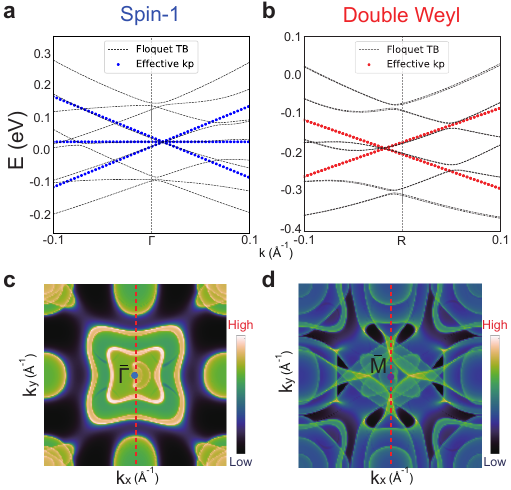}
		\caption{{\bf Chiral light induced momentum shifts for topological fermions in the LH CoSi.} {\bf a,b} The Floquet band structures around $\Gamma$ and R points for the LH CoSi under the radiation of the LCPL without SOC. Blue and red dot lines in (\textbf{a}) and (\textbf{b}) are obtained from the Floquet effective $\mathbf{k}\cdot\mathbf{p}$ model and the black dashed lines are obtained from the diagonalization of the Floquet tight-binding Hamiltonian. The cut-off of the Floquet index is set as $n=\{-1, 0, 1\}$, and the Fermi levels are set as zero. In these calculations, the photon energy is 100~meV and the electric field intensity is $4.4\times10^7$~V/m. {\bf c,d} The projection of the bulk states around $\Gamma$ and R points onto the (001) side surface respectively, where $\bar{\Gamma}$ and $\rm{\bar{M}}$ points are their projected points. The energy levels in (\textbf{c}) and (\textbf{d}) are set at crossing points of the spin-1 excitation and double Weyl fermion, respectively.}
		\label{fig:2}
	\end{figure}
	
	\begin{figure}
		\centering
		\includegraphics[width=0.8\linewidth]{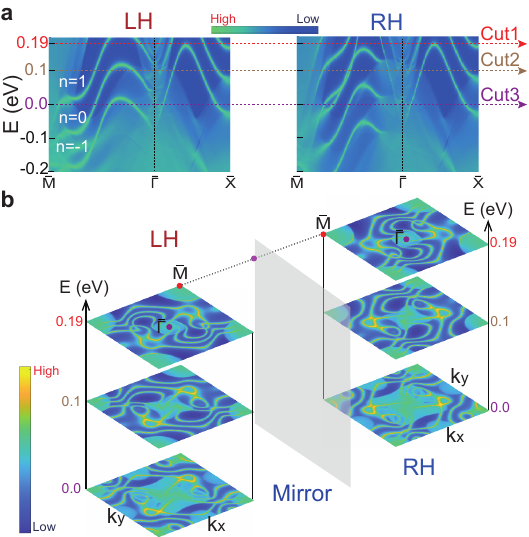}
		\caption{{\bf Surface states in the LH and RH CoSi under chiral Floquet engineering.} {\bf a} The surface Floquet electronic structures along high symmetry lines, which are projected to the (001) surface for the LH (left panel) and RH (right panel) CoSi without SOC under the LCPL radiation. {\bf b} The evolution of the constant-energy contours for Floquet surface states in the LH (left panel) and RH (right panel) CoSi without SOC. Three energy cuts are marked in (\textbf{a}) and the mirror plane is a gray parallelogram.}\label{fig:3}
	\end{figure}

	\begin{figure}[h]
		\centering
		\includegraphics[width=0.8\columnwidth]{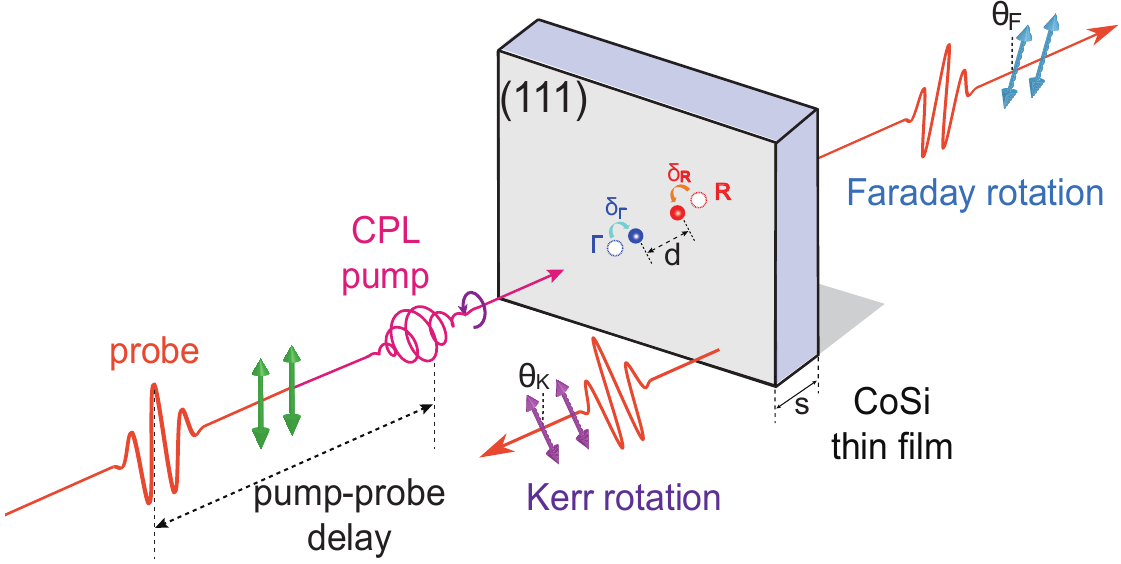}
		\caption{{\bf The setup schematic of the pump-probe Kerr and Faraday experiments on the CoSi thin film sample.} The CPL (pink) are illuminated onto the CoSi thin film. The polarization directions of the reflected and transmitted light are denoted as purple and blue arrows, and the Kerr ($\theta_K$) and Faraday ($\theta_F$) angles of the linearly polarized probe pulse (orange) are measured as a function of pump-probe delay time. }\label{fig:4}
	\end{figure}
	

\subsection{References}
	\begin{thebibliography}{10}
		\expandafter\ifx\csname url\endcsname\relax
		\def\url#1{\texttt{#1}}\fi
		\expandafter\ifx\csname urlprefix\endcsname\relax\def\urlprefix{URL }\fi
		\providecommand{\bibinfo}[2]{#2}
		\providecommand{\eprint}[2][]{\url{#2}}
		
		\bibitem{song2016chiral}
		\bibinfo{author}{Song, J.~C.} \& \bibinfo{author}{Rudner, M.~S.}
		\newblock \bibinfo{title}{Chiral plasmons without magnetic field}.
		\newblock \emph{\bibinfo{journal}{Proc. Natl. Acad. Sci. U.S.A.}}
		\textbf{\bibinfo{volume}{113}}, \bibinfo{pages}{4658--4663}
		(\bibinfo{year}{2016}).
		
		\bibitem{yang2021chiral}
		\bibinfo{author}{Yang, S.-H.}, \bibinfo{author}{Naaman, R.},
		\bibinfo{author}{Paltiel, Y.} \& \bibinfo{author}{Parkin, S.~S.}
		\newblock \bibinfo{title}{Chiral spintronics}.
		\newblock \emph{\bibinfo{journal}{Nat. Rev. Phys.}}
		\textbf{\bibinfo{volume}{3}}, \bibinfo{pages}{328--343}
		(\bibinfo{year}{2021}).
		
		\bibitem{liu2021chirality}
		\bibinfo{author}{Liu, Y.}, \bibinfo{author}{Xiao, J.}, \bibinfo{author}{Koo,
			J.} \& \bibinfo{author}{Yan, B.}
		\newblock \bibinfo{title}{Chirality-driven topological electronic structure of
			{DNA}-like materials}.
		\newblock \emph{\bibinfo{journal}{Nat. Mater.}} \textbf{\bibinfo{volume}{20}},
		\bibinfo{pages}{638--644} (\bibinfo{year}{2021}).
		
		\bibitem{fu2022quantum}
		\bibinfo{author}{Fu, B.}, \bibinfo{author}{Zou, J.-Y.}, \bibinfo{author}{Hu,
			Z.-A.}, \bibinfo{author}{Wang, H.-W.} \& \bibinfo{author}{Shen, S.-Q.}
		\newblock \bibinfo{title}{Quantum anomalous semimetals}.
		\newblock \emph{\bibinfo{journal}{npj Quantum Mater.}}
		\textbf{\bibinfo{volume}{7}}, \bibinfo{pages}{94} (\bibinfo{year}{2022}).
		
		\bibitem{chang2018topological}
		\bibinfo{author}{Chang, G.} \emph{et~al.}
		\newblock \bibinfo{title}{Topological quantum properties of chiral crystals}.
		\newblock \emph{\bibinfo{journal}{Nat. Mater.}} \textbf{\bibinfo{volume}{17}},
		\bibinfo{pages}{978--985} (\bibinfo{year}{2018}).
		
		\bibitem{li2019chiral}
		\bibinfo{author}{Li, H.} \emph{et~al.}
		\newblock \bibinfo{title}{Chiral fermion reversal in chiral crystals}.
		\newblock \emph{\bibinfo{journal}{Nat. Commun.}} \textbf{\bibinfo{volume}{10}},
		\bibinfo{pages}{5505} (\bibinfo{year}{2019}).
		
		\bibitem{li2022chirality}
		\bibinfo{author}{Li, G.} \emph{et~al.}
		\newblock \bibinfo{title}{Chirality locking charge density waves in a chiral
			crystal}.
		\newblock \emph{\bibinfo{journal}{Nat. Commun.}} \textbf{\bibinfo{volume}{13}},
		\bibinfo{pages}{2914} (\bibinfo{year}{2022}).
		
		\bibitem{tang2017multiple}
		\bibinfo{author}{Tang, P.}, \bibinfo{author}{Zhou, Q.} \&
		\bibinfo{author}{Zhang, S.-C.}
		\newblock \bibinfo{title}{Multiple types of topological fermions in transition
			metal silicides}.
		\newblock \emph{\bibinfo{journal}{Phys. Rev. Lett.}}
		\textbf{\bibinfo{volume}{119}}, \bibinfo{pages}{206402}
		(\bibinfo{year}{2017}).
		
		\bibitem{chang2017unconventional}
		\bibinfo{author}{Chang, G.} \emph{et~al.}
		\newblock \bibinfo{title}{Unconventional chiral fermions and large topological
			{F}ermi arcs in {R}h{S}i}.
		\newblock \emph{\bibinfo{journal}{Phys. Rev. Lett.}}
		\textbf{\bibinfo{volume}{119}}, \bibinfo{pages}{206401}
		(\bibinfo{year}{2017}).
		
		\bibitem{ma2017direct}
		\bibinfo{author}{Ma, Q.} \emph{et~al.}
		\newblock \bibinfo{title}{Direct optical detection of {W}eyl fermion chirality
			in a topological semimetal}.
		\newblock \emph{\bibinfo{journal}{Nat. Phys.}} \textbf{\bibinfo{volume}{13}},
		\bibinfo{pages}{842--847} (\bibinfo{year}{2017}).
		
		\bibitem{bradlyn2016beyond}
		\bibinfo{author}{Bradlyn, B.} \emph{et~al.}
		\newblock \bibinfo{title}{Beyond {D}irac and {W}eyl fermions: Unconventional
			quasiparticles in conventional crystals}.
		\newblock \emph{\bibinfo{journal}{Science}} \textbf{\bibinfo{volume}{353}},
		\bibinfo{pages}{aaf5037} (\bibinfo{year}{2016}).
		
		\bibitem{hasan2021weyl}
		\bibinfo{author}{Hasan, M.~Z.} \emph{et~al.}
		\newblock \bibinfo{title}{{W}eyl, {D}irac and high-fold chiral fermions in
			topological quantum matter}.
		\newblock \emph{\bibinfo{journal}{Nat. Rev. Mater.}}
		\textbf{\bibinfo{volume}{6}}, \bibinfo{pages}{784--803}
		(\bibinfo{year}{2021}).
		
		\bibitem{schroter2019chiral}
		\bibinfo{author}{Schr{\"o}ter, N.~B.} \emph{et~al.}
		\newblock \bibinfo{title}{Chiral topological semimetal with multifold band
			crossings and long {F}ermi arcs}.
		\newblock \emph{\bibinfo{journal}{Nat. Phys.}} \textbf{\bibinfo{volume}{15}},
		\bibinfo{pages}{759--765} (\bibinfo{year}{2019}).
		
		\bibitem{rao2019observation}
		\bibinfo{author}{Rao, Z.} \emph{et~al.}
		\newblock \bibinfo{title}{Observation of unconventional chiral fermions with
			long {F}ermi arcs in {C}o{S}i}.
		\newblock \emph{\bibinfo{journal}{Nature}} \textbf{\bibinfo{volume}{567}},
		\bibinfo{pages}{496--499} (\bibinfo{year}{2019}).
		
		\bibitem{sanchez2019topological}
		\bibinfo{author}{Sanchez, D.~S.} \emph{et~al.}
		\newblock \bibinfo{title}{Topological chiral crystals with helicoid-arc quantum
			states}.
		\newblock \emph{\bibinfo{journal}{Nature}} \textbf{\bibinfo{volume}{567}},
		\bibinfo{pages}{500--505} (\bibinfo{year}{2019}).
		
		\bibitem{takane2019observation}
		\bibinfo{author}{Takane, D.} \emph{et~al.}
		\newblock \bibinfo{title}{{Observation of chiral fermions with a large
				topological charge and associated Fermi-arc surface states in CoSi}}.
		\newblock \emph{\bibinfo{journal}{Phys. Rev. Lett.}}
		\textbf{\bibinfo{volume}{122}}, \bibinfo{pages}{076402}
		(\bibinfo{year}{2019}).
		
		\bibitem{schroter2020observation}
		\bibinfo{author}{Schr{\"o}ter, N.~B.} \emph{et~al.}
		\newblock \bibinfo{title}{Observation and control of maximal {C}hern numbers in
			a chiral topological semimetal}.
		\newblock \emph{\bibinfo{journal}{Science}} \textbf{\bibinfo{volume}{369}},
		\bibinfo{pages}{179--183} (\bibinfo{year}{2020}).
		
		\bibitem{de2017quantized}
		\bibinfo{author}{De~Juan, F.}, \bibinfo{author}{Grushin, A.~G.},
		\bibinfo{author}{Morimoto, T.} \& \bibinfo{author}{Moore, J.~E.}
		\newblock \bibinfo{title}{Quantized circular photogalvanic effect in {W}eyl
			semimetals}.
		\newblock \emph{\bibinfo{journal}{Nat. Commun.}} \textbf{\bibinfo{volume}{8}},
		\bibinfo{pages}{15995} (\bibinfo{year}{2017}).
		
		\bibitem{flicker2018chiral}
		\bibinfo{author}{Flicker, F.} \emph{et~al.}
		\newblock \bibinfo{title}{Chiral optical response of multifold fermions}.
		\newblock \emph{\bibinfo{journal}{Phys. Rev. B}} \textbf{\bibinfo{volume}{98}},
		\bibinfo{pages}{155145} (\bibinfo{year}{2018}).
		
		\bibitem{rees2020helicity}
		\bibinfo{author}{Rees, D.} \emph{et~al.}
		\newblock \bibinfo{title}{Helicity-dependent photocurrents in the chiral {W}eyl
			semimetal {R}h{S}i}.
		\newblock \emph{\bibinfo{journal}{Sci. Adv.}} \textbf{\bibinfo{volume}{6}},
		\bibinfo{pages}{eaba0509} (\bibinfo{year}{2020}).
		
		\bibitem{xu2020optical}
		\bibinfo{author}{Xu, B.} \emph{et~al.}
		\newblock \bibinfo{title}{Optical signatures of multifold fermions in the
			chiral topological semimetal {C}o{S}i}.
		\newblock \emph{\bibinfo{journal}{Proc. Natl. Acad. Sci. U.S.A.}}
		\textbf{\bibinfo{volume}{117}}, \bibinfo{pages}{27104--27110}
		(\bibinfo{year}{2020}).
		
		\bibitem{yuan2019quasiparticle}
		\bibinfo{author}{Yuan, Q.-Q.} \emph{et~al.}
		\newblock \bibinfo{title}{Quasiparticle interference evidence of the
			topological {F}ermi arc states in chiral fermionic semimetal {C}o{S}i}.
		\newblock \emph{\bibinfo{journal}{Sci. Adv.}} \textbf{\bibinfo{volume}{5}},
		\bibinfo{pages}{eaaw9485} (\bibinfo{year}{2019}).
		
		\bibitem{sessi2020handedness}
		\bibinfo{author}{Sessi, P.} \emph{et~al.}
		\newblock \bibinfo{title}{Handedness-dependent quasiparticle interference in
			the two enantiomers of the topological chiral semimetal {P}d{G}a}.
		\newblock \emph{\bibinfo{journal}{Nat. Commun.}} \textbf{\bibinfo{volume}{11}},
		\bibinfo{pages}{3507} (\bibinfo{year}{2020}).
		
		\bibitem{lien2023unconventional}
		\bibinfo{author}{Lien, S.-W.} \emph{et~al.}
		\newblock \bibinfo{title}{{Unconventional resistivity scaling in topological
				semimetal CoSi}}.
		\newblock \emph{\bibinfo{journal}{npj Quantum Mater.}}
		\textbf{\bibinfo{volume}{8}}, \bibinfo{pages}{3} (\bibinfo{year}{2023}).
		
		\bibitem{yang2023monopolelike}
		\bibinfo{author}{Yang, Q.} \emph{et~al.}
		\newblock \bibinfo{title}{Monopole-like orbital-momentum locking and the
			induced orbital transport in topological chiral semimetals}.
		\newblock \emph{\bibinfo{journal}{Proc. Natl. Acad. Sci. U.S.A.}}
		\textbf{\bibinfo{volume}{120}}, \bibinfo{pages}{e2305541120}
		(\bibinfo{year}{2023}).
		
		\bibitem{dutta2022collective}
		\bibinfo{author}{Dutta, D.} \emph{et~al.}
		\newblock \bibinfo{title}{Collective plasmonic modes in the chiral multifold
			fermionic material {C}o{S}i}.
		\newblock \emph{\bibinfo{journal}{Phys. Rev. B}}
		\textbf{\bibinfo{volume}{105}}, \bibinfo{pages}{165104}
		(\bibinfo{year}{2022}).
		
		\bibitem{huber2023quantum}
		\bibinfo{author}{Huber, N.} \emph{et~al.}
		\newblock \bibinfo{title}{Quantum oscillations of the quasiparticle lifetime in
			a metal}.
		\newblock \emph{\bibinfo{journal}{Nature}} \textbf{\bibinfo{volume}{621}},
		\bibinfo{pages}{276--281} (\bibinfo{year}{2023}).
		
		\bibitem{guo2022quasi}
		\bibinfo{author}{Guo, C.} \emph{et~al.}
		\newblock \bibinfo{title}{Quasi-symmetry-protected topology in a semi-metal}.
		\newblock \emph{\bibinfo{journal}{Nat. Phys.}} \textbf{\bibinfo{volume}{18}},
		\bibinfo{pages}{813--818} (\bibinfo{year}{2022}).
		
		\bibitem{HsiehDemond2017}
		\bibinfo{author}{Basov, D.~N.}, \bibinfo{author}{Averitt, R.~D.} \&
		\bibinfo{author}{Hsieh, D.}
		\newblock \bibinfo{title}{Towards properties on demand in quantum materials}.
		\newblock \emph{\bibinfo{journal}{Nat. Mater.}} \textbf{\bibinfo{volume}{16}},
		\bibinfo{pages}{1077--1088} (\bibinfo{year}{2017}).
		
		\bibitem{wang2018theoretical}
		\bibinfo{author}{Wang, Y.} \emph{et~al.}
		\newblock \bibinfo{title}{Theoretical understanding of photon spectroscopies in
			correlated materials in and out of equilibrium}.
		\newblock \emph{\bibinfo{journal}{Nat. Rev. Mater.}}
		\textbf{\bibinfo{volume}{3}}, \bibinfo{pages}{312--323}
		(\bibinfo{year}{2018}).
		
		\bibitem{oka2019floquet}
		\bibinfo{author}{Oka, T.} \& \bibinfo{author}{Kitamura, S.}
		\newblock \bibinfo{title}{Floquet engineering of quantum materials}.
		\newblock \emph{\bibinfo{journal}{Annu. Rev. Condens. Matter Phys.}}
		\textbf{\bibinfo{volume}{10}}, \bibinfo{pages}{387--408}
		(\bibinfo{year}{2019}).
		
		\bibitem{DelaTorre2021}
		\bibinfo{author}{de~la Torre, A.} \emph{et~al.}
		\newblock \bibinfo{title}{Colloquium: {N}onthermal pathways to ultrafast
			control in quantum materials}.
		\newblock \emph{\bibinfo{journal}{Rev. Mod. Phys.}}
		\textbf{\bibinfo{volume}{93}}, \bibinfo{pages}{041002}
		(\bibinfo{year}{2021}).
		
		\bibitem{bao2022light}
		\bibinfo{author}{Bao, C.}, \bibinfo{author}{Tang, P.}, \bibinfo{author}{Sun,
			D.} \& \bibinfo{author}{Zhou, S.}
		\newblock \bibinfo{title}{Light-induced emergent phenomena in 2{D} materials
			and topological materials}.
		\newblock \emph{\bibinfo{journal}{Nat. Rev. Phys.}}
		\textbf{\bibinfo{volume}{4}}, \bibinfo{pages}{33--48} (\bibinfo{year}{2022}).
		
		\bibitem{WangYH2013}
		\bibinfo{author}{Wang, Y.}, \bibinfo{author}{Steinberg, H.},
		\bibinfo{author}{Jarillo-Herrero, P.} \& \bibinfo{author}{Gedik, N.}
		\newblock \bibinfo{title}{{Observation of Floquet-Bloch states on the surface
				of a topological insulator}}.
		\newblock \emph{\bibinfo{journal}{Science}} \textbf{\bibinfo{volume}{342}},
		\bibinfo{pages}{453--457} (\bibinfo{year}{2013}).
		
		\bibitem{ito2023build}
		\bibinfo{author}{Ito, S.} \emph{et~al.}
		\newblock \bibinfo{title}{{Build-up and dephasing of Floquet--Bloch bands on
				subcycle timescales}}.
		\newblock \emph{\bibinfo{journal}{Nature}} \textbf{\bibinfo{volume}{616}},
		\bibinfo{pages}{696--701} (\bibinfo{year}{2023}).
		
		\bibitem{mciver2020light}
		\bibinfo{author}{McIver, J.~W.} \emph{et~al.}
		\newblock \bibinfo{title}{{Light-induced anomalous Hall effect in graphene}}.
		\newblock \emph{\bibinfo{journal}{Nat. Phys.}} \textbf{\bibinfo{volume}{16}},
		\bibinfo{pages}{38--41} (\bibinfo{year}{2020}).
		
		\bibitem{sato2019microscopic}
		\bibinfo{author}{Sato, S.} \emph{et~al.}
		\newblock \bibinfo{title}{{Microscopic theory for the light-induced anomalous
				Hall effect in graphene}}.
		\newblock \emph{\bibinfo{journal}{Phys. Rev. B}} \textbf{\bibinfo{volume}{99}},
		\bibinfo{pages}{214302} (\bibinfo{year}{2019}).
		
		\bibitem{murotani2023disentangling}
		\bibinfo{author}{Murotani, Y.} \emph{et~al.}
		\newblock \bibinfo{title}{{Disentangling the competing mechanisms of
				light-induced anomalous Hall conductivity in three-dimensional Dirac
				semimetal}}.
		\newblock \emph{\bibinfo{journal}{Phys. Rev. Lett.}}
		\textbf{\bibinfo{volume}{131}}, \bibinfo{pages}{096901}
		(\bibinfo{year}{2023}).
		
		\bibitem{chan2016chiral}
		\bibinfo{author}{Chan, C.-K.}, \bibinfo{author}{Lee, P.~A.},
		\bibinfo{author}{Burch, K.~S.}, \bibinfo{author}{Han, J.~H.} \&
		\bibinfo{author}{Ran, Y.}
		\newblock \bibinfo{title}{{When chiral photons meet chiral fermions:
				photoinduced anomalous Hall effects in Weyl semimetals}}.
		\newblock \emph{\bibinfo{journal}{Phys. Rev. Lett.}}
		\textbf{\bibinfo{volume}{116}}, \bibinfo{pages}{026805}
		(\bibinfo{year}{2016}).
		
		\bibitem{hubener2017creating}
		\bibinfo{author}{H{\"u}bener, H.}, \bibinfo{author}{Sentef, M.~A.},
		\bibinfo{author}{De~Giovannini, U.}, \bibinfo{author}{Kemper, A.~F.} \&
		\bibinfo{author}{Rubio, A.}
		\newblock \bibinfo{title}{{Creating stable Floquet-{W}eyl semimetals by
				laser-driving of 3D {D}irac materials}}.
		\newblock \emph{\bibinfo{journal}{Nat. Commun.}} \textbf{\bibinfo{volume}{8}},
		\bibinfo{pages}{13940} (\bibinfo{year}{2017}).
		
		\bibitem{LiXiaoShi2019}
		\bibinfo{author}{Li, X.-S.} \emph{et~al.}
		\newblock \bibinfo{title}{Photon-induced {W}eyl half-metal phase and spin
			filter effect from topological {D}irac semimetals}.
		\newblock \emph{\bibinfo{journal}{Phys. Rev. Lett.}}
		\textbf{\bibinfo{volume}{123}}, \bibinfo{pages}{206601}
		(\bibinfo{year}{2019}).
		
		\bibitem{lindner2011floquet}
		\bibinfo{author}{Lindner, N.~H.}, \bibinfo{author}{Refael, G.} \&
		\bibinfo{author}{Galitski, V.}
		\newblock \bibinfo{title}{{Floquet topological insulator in semiconductor
				quantum wells}}.
		\newblock \emph{\bibinfo{journal}{Nat. Phys.}} \textbf{\bibinfo{volume}{7}},
		\bibinfo{pages}{490--495} (\bibinfo{year}{2011}).
		
		\bibitem{Lindner2020}
		\bibinfo{author}{Rudner, M.~S.} \& \bibinfo{author}{Lindner, N.~H.}
		\newblock \bibinfo{title}{Band structure engineering and non-equilibrium
			dynamics in {F}loquet topological insulators}.
		\newblock \emph{\bibinfo{journal}{Nat. Rev. Phys.}}
		\textbf{\bibinfo{volume}{2}}, \bibinfo{pages}{229--244}
		(\bibinfo{year}{2020}).
		
		\bibitem{zhan2023floquet}
		\bibinfo{author}{Zhan, F.} \emph{et~al.}
		\newblock \bibinfo{title}{{Floquet engineering of nonequilibrium
				valley-polarized quantum anomalous Hall Effect with tunable Chern number}}.
		\newblock \emph{\bibinfo{journal}{Nano Lett.}} \textbf{\bibinfo{volume}{23}},
		\bibinfo{pages}{2166--2172} (\bibinfo{year}{2023}).
		
		\bibitem{liu2018photoinduced}
		\bibinfo{author}{Liu, H.}, \bibinfo{author}{Sun, J.-T.},
		\bibinfo{author}{Cheng, C.}, \bibinfo{author}{Liu, F.} \&
		\bibinfo{author}{Meng, S.}
		\newblock \bibinfo{title}{Photoinduced nonequilibrium topological states in
			strained black phosphorus}.
		\newblock \emph{\bibinfo{journal}{Phys. Rev. Lett.}}
		\textbf{\bibinfo{volume}{120}}, \bibinfo{pages}{237403}
		(\bibinfo{year}{2018}).
		
		\bibitem{liu2023floquet}
		\bibinfo{author}{Liu, X.} \emph{et~al.}
		\newblock \bibinfo{title}{Floquet engineering of magnetism in topological
			insulator thin films}.
		\newblock \emph{\bibinfo{journal}{Electron. Struct.}}
		\textbf{\bibinfo{volume}{5}}, \bibinfo{pages}{024002} (\bibinfo{year}{2023}).
		
		\bibitem{zhu2023floquet}
		\bibinfo{author}{Zhu, T.}, \bibinfo{author}{Wang, H.} \&
		\bibinfo{author}{Zhang, H.}
		\newblock \bibinfo{title}{{Floquet engineering of magnetic topological
				insulator MnBi$_2$Te$_4$ films}}.
		\newblock \emph{\bibinfo{journal}{Phys. Rev. B}}
		\textbf{\bibinfo{volume}{107}}, \bibinfo{pages}{085151}
		(\bibinfo{year}{2023}).
		
		\bibitem{oka2009photovoltaic}
		\bibinfo{author}{Oka, T.} \& \bibinfo{author}{Aoki, H.}
		\newblock \bibinfo{title}{{Photovoltaic Hall effect in graphene}}.
		\newblock \emph{\bibinfo{journal}{Phys. Rev. B}} \textbf{\bibinfo{volume}{79}},
		\bibinfo{pages}{081406} (\bibinfo{year}{2009}).
		
		\bibitem{Inoue2010}
		\bibinfo{author}{Inoue, J.-i.} \& \bibinfo{author}{Tanaka, A.}
		\newblock \bibinfo{title}{Photoinduced transition between conventional and
			topological insulators in two-dimensional electronic systems}.
		\newblock \emph{\bibinfo{journal}{Phys. Rev. Lett.}}
		\textbf{\bibinfo{volume}{105}}, \bibinfo{pages}{017401}
		(\bibinfo{year}{2010}).
		
		\bibitem{sentef2015theory}
		\bibinfo{author}{Sentef, M.} \emph{et~al.}
		\newblock \bibinfo{title}{{Theory of Floquet band formation and local
				pseudospin textures in pump-probe photoemission of graphene}}.
		\newblock \emph{\bibinfo{journal}{Nat. Commun.}} \textbf{\bibinfo{volume}{6}},
		\bibinfo{pages}{7047} (\bibinfo{year}{2015}).
		
		\bibitem{yan2016tunable}
		\bibinfo{author}{Yan, Z.} \& \bibinfo{author}{Wang, Z.}
		\newblock \bibinfo{title}{Tunable {W}eyl points in periodically driven nodal
			line semimetals}.
		\newblock \emph{\bibinfo{journal}{Phys. Rev. Lett.}}
		\textbf{\bibinfo{volume}{117}}, \bibinfo{pages}{087402}
		(\bibinfo{year}{2016}).
		
		\bibitem{zhang2016theory}
		\bibinfo{author}{Zhang, X.-X.}, \bibinfo{author}{Ong, T.~T.} \&
		\bibinfo{author}{Nagaosa, N.}
		\newblock \bibinfo{title}{{Theory of photoinduced Floquet Weyl semimetal
				phases}}.
		\newblock \emph{\bibinfo{journal}{Phys. Rev. B}} \textbf{\bibinfo{volume}{94}},
		\bibinfo{pages}{235137} (\bibinfo{year}{2016}).
		
		\bibitem{chen2018floquet}
		\bibinfo{author}{Chen, Q.}, \bibinfo{author}{Du, L.} \& \bibinfo{author}{Fiete,
			G.~A.}
		\newblock \bibinfo{title}{Floquet band structure of a semi-{D}irac system}.
		\newblock \emph{\bibinfo{journal}{Phys. Rev. B}} \textbf{\bibinfo{volume}{97}},
		\bibinfo{pages}{035422} (\bibinfo{year}{2018}).
		
		\bibitem{trevisan2022bicircular}
		\bibinfo{author}{Trevisan, T.~V.}, \bibinfo{author}{Arribi, P.~V.},
		\bibinfo{author}{Heinonen, O.}, \bibinfo{author}{Slager, R.-J.} \&
		\bibinfo{author}{Orth, P.~P.}
		\newblock \bibinfo{title}{{Bicircular light Floquet engineering of magnetic
				symmetry and topology and its application to the Dirac semimetal
				Cd$_3$As$_2$}}.
		\newblock \emph{\bibinfo{journal}{Phys. Rev. Lett.}}
		\textbf{\bibinfo{volume}{128}}, \bibinfo{pages}{066602}
		(\bibinfo{year}{2022}).
		
		\bibitem{sie2019ultrafast}
		\bibinfo{author}{Sie, E.~J.} \emph{et~al.}
		\newblock \bibinfo{title}{{An ultrafast symmetry switch in a Weyl semimetal}}.
		\newblock \emph{\bibinfo{journal}{Nature}} \textbf{\bibinfo{volume}{565}},
		\bibinfo{pages}{61--66} (\bibinfo{year}{2019}).
		
		\bibitem{kitayama2021predicted}
		\bibinfo{author}{Kitayama, K.}, \bibinfo{author}{Mochizuki, M.},
		\bibinfo{author}{Tanaka, Y.} \& \bibinfo{author}{Ogata, M.}
		\newblock \bibinfo{title}{{Predicted photoinduced pair annihilation of emergent
				magnetic charges in the organic salt $\alpha$-(BEDT-TTF)$_2$I$_3$ irradiated
				by linearly polarized light}}.
		\newblock \emph{\bibinfo{journal}{Phys. Rev. B}}
		\textbf{\bibinfo{volume}{104}}, \bibinfo{pages}{075127}
		(\bibinfo{year}{2021}).
		
		\bibitem{neufeld2023band}
		\bibinfo{author}{Neufeld, O.}, \bibinfo{author}{H{\"u}bener, H.},
		\bibinfo{author}{Jotzu, G.}, \bibinfo{author}{De~Giovannini, U.} \&
		\bibinfo{author}{Rubio, A.}
		\newblock \bibinfo{title}{{Band nonlinearity-enabled manipulation of Dirac
				nodes, Weyl cones, and valleytronics with intense linearly polarized light}}.
		\newblock \emph{\bibinfo{journal}{Nano Lett.}} \textbf{\bibinfo{volume}{23}},
		\bibinfo{pages}{7568--7575} (\bibinfo{year}{2023}).
		
		\bibitem{yoshikawa2022light}
		\bibinfo{author}{Yoshikawa, N.} \emph{et~al.}
		\newblock \bibinfo{title}{{Light-induced chiral gauge field in a massive 3D
				Dirac electron system}}.
		\newblock \emph{\bibinfo{journal}{arXiv preprint arXiv:2209.11932}}
		(\bibinfo{year}{2022}).
		
		\bibitem{Mikami2016Brillouin}
		\bibinfo{author}{Mikami, T.} \emph{et~al.}
		\newblock \bibinfo{title}{{Brillouin-Wigner theory for high-frequency expansion
				in periodically driven systems: Application to Floquet topological
				insulators}}.
		\newblock \emph{\bibinfo{journal}{Phys. Rev. B}} \textbf{\bibinfo{volume}{93}},
		\bibinfo{pages}{144307} (\bibinfo{year}{2016}).
		
		\bibitem{mahmood2016selective}
		\bibinfo{author}{Mahmood, F.} \emph{et~al.}
		\newblock \bibinfo{title}{{Selective scattering between Floquet--Bloch and
				Volkov states in a topological insulator}}.
		\newblock \emph{\bibinfo{journal}{Nat. Phys.}} \textbf{\bibinfo{volume}{12}},
		\bibinfo{pages}{306--310} (\bibinfo{year}{2016}).
		
		\bibitem{Wang2014Stark}
		\bibinfo{author}{Kim, J.} \emph{et~al.}
		\newblock \bibinfo{title}{{Ultrafast generation of pseudo-magnetic field for
				valley excitons in WSe$_2$ monolayers}}.
		\newblock \emph{\bibinfo{journal}{Science}} \textbf{\bibinfo{volume}{346}},
		\bibinfo{pages}{1205--1208} (\bibinfo{year}{2014}).
		
		\bibitem{sie2017large}
		\bibinfo{author}{Sie, E.~J.} \emph{et~al.}
		\newblock \bibinfo{title}{{Large, valley-exclusive Bloch-Siegert shift in
				monolayer WS$_2$}}.
		\newblock \emph{\bibinfo{journal}{Science}} \textbf{\bibinfo{volume}{355}},
		\bibinfo{pages}{1066--1069} (\bibinfo{year}{2017}).
		
		\bibitem{Aeschlimann2021Survival}
		\bibinfo{author}{Aeschlimann, S.} \emph{et~al.}
		\newblock \bibinfo{title}{{Survival of Floquet-Bloch states in the presence of
				scattering}}.
		\newblock \emph{\bibinfo{journal}{Nano Lett.}} \textbf{\bibinfo{volume}{21}},
		\bibinfo{pages}{5028--5035} (\bibinfo{year}{2021}).
		
		\bibitem{zhou2023pseudospin}
		\bibinfo{author}{Zhou, S.} \emph{et~al.}
		\newblock \bibinfo{title}{{Pseudospin-selective Floquet band engineering in
				black phosphorus}}.
		\newblock \emph{\bibinfo{journal}{Nature}} \textbf{\bibinfo{volume}{614}},
		\bibinfo{pages}{75--80} (\bibinfo{year}{2023}).
		
		\bibitem{zhou2023Floquet}
		\bibinfo{author}{Zhou, S.} \emph{et~al.}
		\newblock \bibinfo{title}{Floquet engineering of black phosphorus upon
			below-gap pumping}.
		\newblock \emph{\bibinfo{journal}{Phys. Rev. Lett.}}
		\textbf{\bibinfo{volume}{131}}, \bibinfo{pages}{116401}
		(\bibinfo{year}{2023}).
		
		\bibitem{merboldt2024observation}
		\bibinfo{author}{Merboldt, M.} \emph{et~al.}
		\newblock \bibinfo{title}{{Observation of Floquet states in graphene}}.
		\newblock \emph{\bibinfo{journal}{arxiv preprint,arXiv.2404.12791}}
		(\bibinfo{year}{2024}).
		
		\bibitem{choi2024direct}
		\bibinfo{author}{Choi, D.} \emph{et~al.}
		\newblock \bibinfo{title}{{Direct observation of Floquet-Bloch states in
				monolayer graphene}}.
		\newblock \emph{\bibinfo{journal}{arxiv preprint,arXiv.2404.14392}}
		(\bibinfo{year}{2024}).
		
		\bibitem{sato2020floquet}
		\bibinfo{author}{Sato, S.} \emph{et~al.}
		\newblock \bibinfo{title}{Floquet states in dissipative open quantum systems}.
		\newblock \emph{\bibinfo{journal}{J. Phys. B}} \textbf{\bibinfo{volume}{53}},
		\bibinfo{pages}{225601} (\bibinfo{year}{2020}).
		
		\bibitem{xu2019crystal}
		\bibinfo{author}{Xu, X.} \emph{et~al.}
		\newblock \bibinfo{title}{Crystal growth and quantum oscillations in the
			topological chiral semimetal {C}o{S}i}.
		\newblock \emph{\bibinfo{journal}{Phys. Rev. B}}
		\textbf{\bibinfo{volume}{100}}, \bibinfo{pages}{045104}
		(\bibinfo{year}{2019}).
		
		\bibitem{chen2020observing}
		\bibinfo{author}{Chen, Y.}, \bibinfo{author}{Wang, Y.},
		\bibinfo{author}{Claassen, M.}, \bibinfo{author}{Moritz, B.} \&
		\bibinfo{author}{Devereaux, T.~P.}
		\newblock \bibinfo{title}{{Observing photo-induced chiral edge states of
				graphene nanoribbons in pump-probe spectroscopies}}.
		\newblock \emph{\bibinfo{journal}{npj Quantum Mater.}}
		\textbf{\bibinfo{volume}{5}}, \bibinfo{pages}{84} (\bibinfo{year}{2020}).
		
		\bibitem{cheong2023trompe}
		\bibinfo{author}{Cheong, S.-W.} \& \bibinfo{author}{Huang, F.-T.}
		\newblock \bibinfo{title}{{Trompe L’oeil Ferromagnetism—magnetic point
				group analysis}}.
		\newblock \emph{\bibinfo{journal}{npj Quantum Mater.}}
		\textbf{\bibinfo{volume}{8}}, \bibinfo{pages}{73} (\bibinfo{year}{2023}).
		
		\bibitem{yang2011quantum}
		\bibinfo{author}{Yang, K.-Y.}, \bibinfo{author}{Lu, Y.-M.} \&
		\bibinfo{author}{Ran, Y.}
		\newblock \bibinfo{title}{{Quantum Hall effects in a Weyl semimetal: Possible
				application in pyrochlore iridates}}.
		\newblock \emph{\bibinfo{journal}{Phys. Rev. B}} \textbf{\bibinfo{volume}{84}},
		\bibinfo{pages}{075129} (\bibinfo{year}{2011}).
		
		\bibitem{hirai2023anomalous}
		\bibinfo{author}{Hirai, Y.} \emph{et~al.}
		\newblock \bibinfo{title}{{Anomalous Hall effect of light-driven
				three-dimensional Dirac electrons in bismuth}}.
		\newblock \emph{\bibinfo{journal}{arXiv preprint arXiv:2301.06072}}
		(\bibinfo{year}{2023}).
		
		\bibitem{kahn1969ultraviolet}
		\bibinfo{author}{Kahn, F.~J.}, \bibinfo{author}{Pershan, P.} \&
		\bibinfo{author}{Remeika, J.}
		\newblock \bibinfo{title}{Ultraviolet magneto-optical properties of
			single-crystal orthoferrites, garnets, and other ferric oxide compounds}.
		\newblock \emph{\bibinfo{journal}{Phys. Rev.}} \textbf{\bibinfo{volume}{186}},
		\bibinfo{pages}{891} (\bibinfo{year}{1969}).
		
		\bibitem{zhang2022chirality}
		\bibinfo{author}{Zhang, Y.} \emph{et~al.}
		\newblock \bibinfo{title}{Chirality logic gates}.
		\newblock \emph{\bibinfo{journal}{Sci. Adv.}} \textbf{\bibinfo{volume}{8}},
		\bibinfo{pages}{eabq8246} (\bibinfo{year}{2022}).
		
		\bibitem{hubener2021engineering}
		\bibinfo{author}{H{\"u}bener, H.} \emph{et~al.}
		\newblock \bibinfo{title}{Engineering quantum materials with chiral optical
			cavities}.
		\newblock \emph{\bibinfo{journal}{Nat. Mater.}} \textbf{\bibinfo{volume}{20}},
		\bibinfo{pages}{438--442} (\bibinfo{year}{2021}).
		
		\bibitem{kresse1996efficient}
		\bibinfo{author}{Kresse, G.} \& \bibinfo{author}{Furthm\"uller, J.}
		\newblock \bibinfo{title}{Efficient iterative schemes for ab initio
			total-energy calculations using a plane-wave basis set}.
		\newblock \emph{\bibinfo{journal}{Phys. Rev. B}} \textbf{\bibinfo{volume}{54}},
		\bibinfo{pages}{11169--11186} (\bibinfo{year}{1996}).
		
		\bibitem{perdew1996generalized}
		\bibinfo{author}{Perdew, J.~P.}, \bibinfo{author}{Burke, K.} \&
		\bibinfo{author}{Ernzerhof, M.}
		\newblock \bibinfo{title}{Generalized gradient approximation made simple}.
		\newblock \emph{\bibinfo{journal}{Phys. Rev. Lett.}}
		\textbf{\bibinfo{volume}{77}}, \bibinfo{pages}{3865--3868}
		(\bibinfo{year}{1996}).
		
		\bibitem{kresse1999ultrasoft}
		\bibinfo{author}{Kresse, G.} \& \bibinfo{author}{Joubert, D.}
		\newblock \bibinfo{title}{From ultrasoft pseudopotentials to the projector
			augmented-wave method}.
		\newblock \emph{\bibinfo{journal}{Phys. Rev. B}} \textbf{\bibinfo{volume}{59}},
		\bibinfo{pages}{1758--1775} (\bibinfo{year}{1999}).
		
		\bibitem{mostofi2008wannier90}
		\bibinfo{author}{Mostofi, A.~A.} \emph{et~al.}
		\newblock \bibinfo{title}{{Wannier90: A tool for obtaining maximally-localised
				Wannier functions}}.
		\newblock \emph{\bibinfo{journal}{Computer Phys. Comm.}}
		\textbf{\bibinfo{volume}{178}}, \bibinfo{pages}{685--699}
		(\bibinfo{year}{2008}).
		
		\bibitem{Sancho1985Highly}
		\bibinfo{author}{Sancho, M.~L.}, \bibinfo{author}{Sancho, J.~L.},
		\bibinfo{author}{Sancho, J.~L.} \& \bibinfo{author}{Rubio, J.}
		\newblock \bibinfo{title}{{Highly convergent schemes for the calculation of
				bulk and surface Green functions}}.
		\newblock \emph{\bibinfo{journal}{J. Phys. F}} \textbf{\bibinfo{volume}{15}},
		\bibinfo{pages}{851} (\bibinfo{year}{1985}).
		
		\bibitem{wu2018wanniertools}
		\bibinfo{author}{Wu, Q.}, \bibinfo{author}{Zhang, S.}, \bibinfo{author}{Song,
			H.-F.}, \bibinfo{author}{Troyer, M.} \& \bibinfo{author}{Soluyanov, A.~A.}
		\newblock \bibinfo{title}{{WannierTools: An open-source software package for
				novel topological materials}}.
		\newblock \emph{\bibinfo{journal}{Comp. Phys. Commun.}}
		\textbf{\bibinfo{volume}{224}}, \bibinfo{pages}{405--416}
		(\bibinfo{year}{2018}).
		
	\end{thebibliography}
\end{document}